**Automated Synthesis of Steady-State Continuous Processes using Reinforcement Learning**

Quirin Göttl[a,*], Dominik G. Grimm[b,c,d], Jakob Burger[a],

**Abstract**

Automated flowsheet synthesis is an important field in computer-aided process engineering. The present work demonstrates how reinforcement learning can be used for automated flowsheet synthesis without any heuristics of prior knowledge of conceptual design. The environment consists of a steady-state flowsheet simulator that contains all physical knowledge. An agent is trained to take discrete actions and sequentially built up flowsheets that solve a given process problem. A novel method named SynGameZero is developed to ensure good exploration schemes in the complex problem. Therein, flowsheet synthesis is modelled as a game of two competing players. The agent plays this game against itself during training and consists of an artificial neural network and a tree search for forward planning. The method is applied successfully to a reaction-distillation process in a quaternary system.

**Keywords**

**Author affiliations**

[a]Technical University of Munich, Campus Straubing for Biotechnology and Sustainability, Laboratory of Chemical Process Engineering, Schulgasse 16, 94315 Straubing, Germany

[b]Technical University of Munich, Campus Straubing for Biotechnology and Sustainability, Bioinformatics, Schulgasse 22, 94315 Straubing, Germany

[c]Weihenstephan-Triesdorf University of Applied Sciences, Petersgasse 18, 94315 Straubing, Germany

[d]Technical University of Munich, Department of Informatics, Bolzmannstr. 3, 85748 Garching, Germany

E-Mail corresponding author: quirin.goettl@tum.de

## 1      Introduction

In chemical engineering, process synthesis can be defined as the act, where one invents the structure and operating levels for a new chemical manufacturing process 1.. Computer-aided process synthesis has been an important field of chemical engineering for decades 2.. There exists a vast amount of methods in computer-aided process synthesis, in which the roles of human and computer are quite different and vary in their proportions. On one end of the spectrum, humans invent flowsheets, provide mechanistic models of apparatus and physico-chemical properties, and employ computers solely in simulations to evaluate and check the invented designs. On the other end of the spectrum, there is automated flowsheet synthesis, which we call rather human-aided process synthesis by a computer. Therein, the structure of the process and operating levels are chosen autonomously by the computer based on input by the human (typically a problem statement and the physico-chemical property data).

Siirola 3. classified automated flowsheet synthesis into three categories: superstructure optimization, evolutionary modification and systematic generation. In superstructure optimization, a large flowsheet structure (the superstructure) is set up in a way, so that a large set of process alternatives can be obtained by removing parts of that structure 4., 5.. An objective function or cost function is defined and the optimal configuration for the flowsheet is determined by an optimization algorithm that uses decision variables to remove parts of the superstructure. Evolutionary modification works as follows: A process flowsheet is devised (by any method at hand), analyzed and changed in one or more ways repeatedly to improve it. The changes are continued until no further improvement in the flowsheet can be made 6.. Systematic generation creates a flowsheet sequentially by adding process units. The decision process is usually based on heuristics, which are based on prior knowledge. Alternatively, it is possible to derive heuristics, by comparing many flowsheets systematically with the help of a computer 7.. Prominent examples of the systematic generation approach are the expert systems 8., 9.. Sometimes two of the three categories are combined in hybrid synthesis methods 10., 11.. For further reading, concerning the state-of-the-art of automated process synthesis, we refer to current review articles 12., 13..

In the present work, a novel machine-learning (ML) based method for automated process synthesis in the category systematic generation is introduced. As ML and artificial intelligence (AI) are rapidly expanding fields, a lot of research focuses on applying these kind of techniques in computer-aided process engineering 14.-18.. In the area of process synthesis, ML is for example applied to create surrogate models for reducing computational time in simulation and optimization 19.-21.. AI offers however more potential, as stated by Dimiduk et al. 16.: "Or, how can one best apply the newest advances in ML and AI to improve MPSE [materials, processes, and structures engineering] results? Speculating still further, why are there no emerging AI-based engineering design systems that recognize component features, attributes, or intended performance to make recommendations about directions for final design, manufacturing processes, and materials selections or developments?" The type of ML techniques, that could address these kind of problems, seems to be reinforcement learning (RL). The objective of RL is to teach an agent, which could for example consist of an artificial neural network (ANN), to master a given task through repeated interactions with its environment 22., 23.. RL is already applied in process engineering, however almost exclusively in process control 24.. Among rare exceptions are Zhou et al. 25., who employed RL to set experimental conditions for the optimization of chemical reactions. Khan and Lapkin 26. used a RL approach to identify promising processing routes in hydrogen production.

Outside process engineering, many authors have proven that RL serves as powerful tool to master difficult problems like winning the board games of Go and Chess by training an agent only through self-play and RL 27., 28.. In the present work, similar techniques are employed to train an agent to come up with good process flowsheets on its own. It solves process synthesis problems using systematic generation and adds process units sequentially and in a constructive way to a flowsheet. The agent is trained without any prior knowledge or heuristics. Through repeated process simulation during the training phase of RL, the agent develops artificial process engineering intuition. The present work is structured as follows: The problem framework and a basic process engineering example problem are defined and explained first. A novel RL method, the agent's structure and the training procedure are explained afterwards in detail. In the Results section, it is shown that the basic process engineering problem is solved by the method, proving the concept to work.

## 2 Experimental

### 2.1 RL framework

The general RL framework for flowsheet synthesis for an agent that has zero prior knowledge is shown in Figure 1. The flowsheet is set up and evaluated in the environment. For reasons of time, cost and safety, the environment is not a real chemical site, but a process simulator (here: a steady-state process simulator). The environment is partially observable: the agent is able to observe the sate $s$ comprising the flowsheet connectivity (which process unit is connected to which) and the stream table that results from the process simulation. Internal states of the process units (e.g. the temperature on some stage of some distillation column) are not part of that state and thus not observable for the agent. The possible actions that the agent can perform on the environment are adding new process units to the flowsheet, setting operational parameters of these units if applicable, adding recycles, or terminating the flowsheet synthesis. As feedback, the agent obtains a reward by the environment, which is generally an improvement in some cost function that is evaluated in the simulator (e.g. net present value of the process).

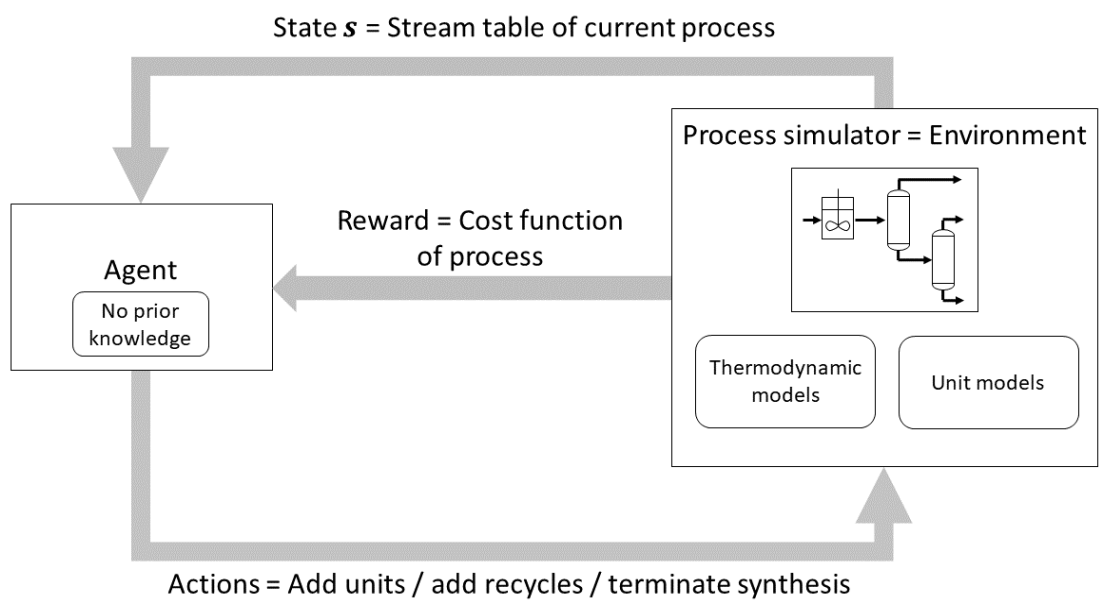

**Figure 1.** Scheme of the RL framework for flowsheet synthesis using only discrete decisions without prior knowledge.

Zero prior knowledge of the agent means that it does not know any property data, thermodynamic model or process unit model *a priori*. Further, neither process engineering heuristics nor any other rule-based schemes are available. Thus, the agent initially takes entirely random actions on the environment. Gradually, it learns better decision policies based on the reward it obtains from the environment. Of course, substantial amounts of physical knowledge have to be supplied somehow to the overall framework. This happens exclusively in the environment: models of thermodynamic properties and the process units have to be supplied by the (human) engineer. Great attention has to be paid to this step, because the agent might exploit every "loophole" in the models. Think of an azeotrope that is not modeled properly in the thermodynamics. The agent will eventually learn to select a purely distillation-based separation if it looks economic, even though it is not feasible in practice due to the azeotrope.

The described framework is quite general. Different types of simulators could be used, as well as different types of agents with various RL methods. In the present work, a subclass of problems is considered: flowsheet synthesis using only discrete decisions without placing recycle streams. This means that the agent solely decides on placing process units one after another, and if necessary on their discrete task. For example, the agent can decide to place a distillation column with the task to separate two components by a sharp split. It does not specify any continuous operation parameter of the unit, as these are specified by its task (the conversion of task to continuous parameter is done by the environment automatically). The reason for this limitation is two-fold. On the one hand, a robust simulation environment can be defined (cf. below). On the other hand, the used RL methods are native in discrete decision spaces. They would require computationally expensive extensions for continuous decision spaces. Despite the limitation to discrete decisions, interesting design problems can be considered as will be described below.

## 2.2 Environment of the case studies

We chose a steady-state flowsheet simulation as the environment. Starting point is one or more feed stream(s) specified by a human engineer. In its actions, the agent is allowed to place a process unit to any open stream (a stream that has no destination yet). After each action, the flowsheet is simulated and a stream table is determined. For the proof of concept in the present work, we kept the environment simple. The processes operate in a model system of four compounds A, B, C and D. The system is zeotropic, thus can be separated by distillation only. If the mixture is fed to a reactor filled with catalyst, the following reaction is observed:

$$A + B \rightarrow C + D. \tag{1}$$

The possible actions of the agent are grouped as follows:

a) Place a distillation column with a perfectly sharp split. The boiling order is ABCD. Thus, the agent has three discrete options called D1 (split A - BCD), D2 (split AB - CD) and D3 (split ABC - D).
b) Place a reactor (denoted as action R). The reactor is a continuous stirred tank reactor with the conversion of A given by a kinetic of first order in A and B:

$$\dot{n}_A^{in} - \dot{n}_A^{out} = 5\frac{\text{kmol}}{\text{hr}} x_A^{out} x_B^{out}. \qquad (2)$$

Therein, "out" and "in" specify quantities at the reactor outlet and inlet, respectively. The variables $\dot{n}_i$ and $x_i$ denote component $i$'s molar flow rate and mole fraction, respectively. The conversion of the other components B, C and D is calculated by the stoichiometry of Reaction (1).
c) Place a mixer for mixing two streams. This action is denoted as M.
d) Terminate the flowsheet synthesis by action T.

The actions D1, D2, D3 and R can be applied to any single open stream, whereas M requires two open streams as input. If more than two streams have to be mixed, the agent could select multiple mixers in a cascade. Implications for more complex problems (more components, detailed apparatus models, complex thermodynamics) are given in the Discussion section.

The net present value is used to evaluate the obtained processes. Since the degree of detail in its calculation is not relevant for the presented methodology and the process models are rather basic, a rather simple scheme is used to calculate the net present value:

$$\text{NPV} = \sum_{u \in U} I_u + (10a) \sum_{o \in O} c_o. \qquad (3)$$

It combines the investment costs $I_u$ of every unit $u$ with the yearly operational cash flows $c_o$ multiplied with 10 years (a factor that lumps the period of depreciation and the interest rates). The investment costs of the units are assumed flat and independent of size and operation parameters, as these quantities are not provided by the model. The yearly operational cash flows $c_o$ consider only cost and revenues from all open material streams $o$ leaving the process. Further operational costs of the units (e.g. steam cost for the distillation) are neglected for simplicity. The cash flows of the open streams are calculated as follows. If a stream contains a pure component $i$:

$$c_o = \dot{n}_i \cdot p_i \cdot 8000 \frac{\text{hr}}{\text{a}}, \qquad (4)$$

where $\dot{n}_i$ is its molar flowrate in kmol/hr and $p_i$ is the price of component $i$. If an open stream is not pure then its yearly cash flow is:

$$c_o = \sum_{i \in \{A,B,C,D\}} \dot{n}_i \cdot \min(p_i, 0) \cdot 8000 \frac{\text{hr}}{\text{a}}. \qquad (5)$$

The minimum function ensures that the cash flow of mixed streams is never positive. If the stream contains a compound of negative price $p_i$ (e.g. a hazardous compound for which disposal has to be paid), then the cash flow becomes negative. The values/costs of the feed stream(s) are not considered explicitly in the formulas, as they are constant and therefore have no influence on finding the optimal process. However, the agent may select the trivial process of placing no process unit at all. In this case, any feed is an open stream leaving the process and is included in the determination of the net present value. In the examples of the present work, two different cases of the cost parameters $I_u$ and $p_i$ are discussed to demonstrate the interchangeability of the cost function. The used parameters are listed in Table 1 in the Results section.

### 2.3 SynGameZero method and agent structure

**Preliminary Analysis.** There are numerous methods for solving the RL problem defined in the framework of Figure 1. Let us start with some general analysis of the problem. The observed state of the environment consists of a list of process units in the flowsheet, their connectivity and a stream table as a result of the process simulation. This observed state fulfils the Markov property, i.e. it is sufficient for solving the flowsheet simulation and the solution is independent from the history of state. This means that future decisions of an optimal agent could be based on the presently observed state alone. The environment behaves fully deterministic, i.e. starting from a certain state and applying a certain action leads always to the same successor state. Thus, an agent can reliably use look-ahead planning methods to evaluate a sequence of actions. It is not trivial to reward the agent after every placement of a new process unit with an immediate and constructive reward. For example, think of a multi-step separation sequence that only works successfully after a recycle has been closed. After placing the first unit of the sequence, a constructive reward is hard to determine. Further, if the agent "forgets" to close the recycle and terminates the flowsheet synthesis, the process simply does not work. Which was the unit that caused the process ultimately to fail? In this case, it is not even possible to reward or punish individual actions constructively after the flowsheet synthesis has been completed. Consequently, we suggest not rewarding every single action, but rather looking only at the final flowsheet and using its value (e.g. net present value) as the reward for the agent. This means that the flowsheet always has to be finished before rewards can be obtained during training of the agent. The observed state contains continuous variables (e.g. the concentration of some compound in some stream), thus there is an infinite number of possible states. This makes the use of look-up tables (i.e. if state is exactly equal to ... then do ...) ineffective. Instead, functions are used by the agent to calculate actions and other quantities from the observed state. Here, ANNs can be used as function approximators. The parameters of the ANNs are learnt during the training phase. Two functions are used by the agent of the present work: a policy function and a value function. The policy function (variable $\pi$) outputs suggestions for the next actions. The value function (variable $v$) outputs the estimated state value that is the expected reward of the final flowsheet when following the policy starting from the present observed state. To exploit both functions, policy-based learning in an actor-critic setup 22., 23. is usually used.

Without going into details, we have naively tried out a setup in which the policy was a vector with a probability distribution for selecting one from all allowed actions for the present state. The value function $v$ was the net present value of the final flowsheet to be expected when following that policy. Using an actor-critic method, we have tried to learn an optimal policy. $v$ was employed as a baseline for the critic. We have quickly discovered that this approach is not constructive. It suffers from fast convergence to local optima, e.g. it produces flowsheets that are only better than highly similar ones. This result is not surprising given the characteristics of the flowsheet problem. As there is no information for the agent on the maximum possible reward, it is not able to determine, whether the learnt policy is a global or a local optimum. Further, flowsheet synthesis is a complex task where one has to think several steps ahead, almost comparable with playing a strategic game like chess where certain moves may pay off only many moves later. In order to suggest breakthrough process units (winning moves in chess), good exploration schemes are necessary during training to avoid following only beaten tracks. Therefore, the present work follows a more sophisticated approach to RL for flowsheet synthesis: we embed the flowsheet synthesis into a competitive two-player game leading to an advantageous learning performance. We call this method *SynGameZero* (Flowsheet Synthesis in a Game Environment with Zero Knowledge) and describe it systematically in the following. The method has a couple of tuning parameters. The values for these parameters are given later in the Results section in Table 2 along the application examples.

**State representation.** The state of a flowsheet is stored into the flowsheet matrix $F$. The construction of $F$ from the simulation results of the environment is explained along Figure 2. Every stream in the flowsheet refers to one row of $F$. $F$ has a fixed number $N_{matrix}$ of rows and if there are less streams in the flowsheet, the remaining rows are filled with zeros ($\mathbf{0}_{1 \times N}$ in Figure 2 refers to a row vector of $N$ zero entries). The number $N_{matrix}$ is an upper limit for the size of the flowsheet (i.e. the number of streams). If the matrix is full then the process synthesis will end automatically. This limitation is due to technical reasons. In practice $N_{matrix}$ has to be chosen large enough to accommodate the optimal flowsheet comfortably. Every row of $F$ is composed of a set of row vectors that are explained along the first row in Figure 2 for stream 1 (reactor input) of the shown flowsheet. $v_1$ contains the molar fractions of all compounds followed by the total molar flow rate of stream 1. The vector $u_1$ specifies the process unit at the streams destination. In the present case study, it has $N_{unit} = 4 + N_{matrix}$ entries. The first four entries refer to distillation splits D1, D2, D3 and reactor R, respectively. The last $N_{matrix}$ entries are relevant if the stream's destination is a mixer. The entry for the corresponding unit is set to 1 and all other entries are set to 0. In case of the mixer, the $(4 + k)$th entry is set to one, indicating that the other stream to the mixer is stream $k$. If no unit is connected to stream $i$, then $u_i$ is set to $\mathbf{0}_{1 \times N_{unit}}$. For example, $u_1$ in Figure 2 indicates that stream 1's destination is a reactor R. The last four vectors of each row contain information on the subsequent streams that leave the process unit and the streams destination. Let us say these are streams $m$ and $n$. The first and the third of the four vectors are copies of $v_m$ and $v_n$, respectively. The second and fourth vector are pointers to the numbers $m$ and $n$, respectively. They are vectors with $N_{matrix}$ entries, all of them 0 but the $m$-th or $n$-th entry, respectively, which are 1. If there is no destination process unit (see e.g. streams 3 or 4) or the process unit has only one output stream (see e.g. stream 1), then all four or the latter two vectors are filled with zeros, respectively. In the present work, only units with up to two output streams are considered, but of course it would be possible to extend the flowsheet matrix $F$ to represent units with more output streams by increasing the length of each line.

There are many other possibilities for state representation including much more compact ones without redundant information. However, the representation above offers a couple of features that we believe are advantageous for RL: Discrete variables describing the flowsheet structure (such as type of process units and stream numbers) are encoded as a single 1 in a vector of zeros rather than in single integer scalars. This method avoids that two different flowsheet structures are falsely considered similar by the function approximation that takes $\boldsymbol{F}$ as input. By repeating the stream information of unit outlets, the states of the input and output streams of a unit are related to each other. This captures that these streams are also closely linked in the environment via the unit's performance.

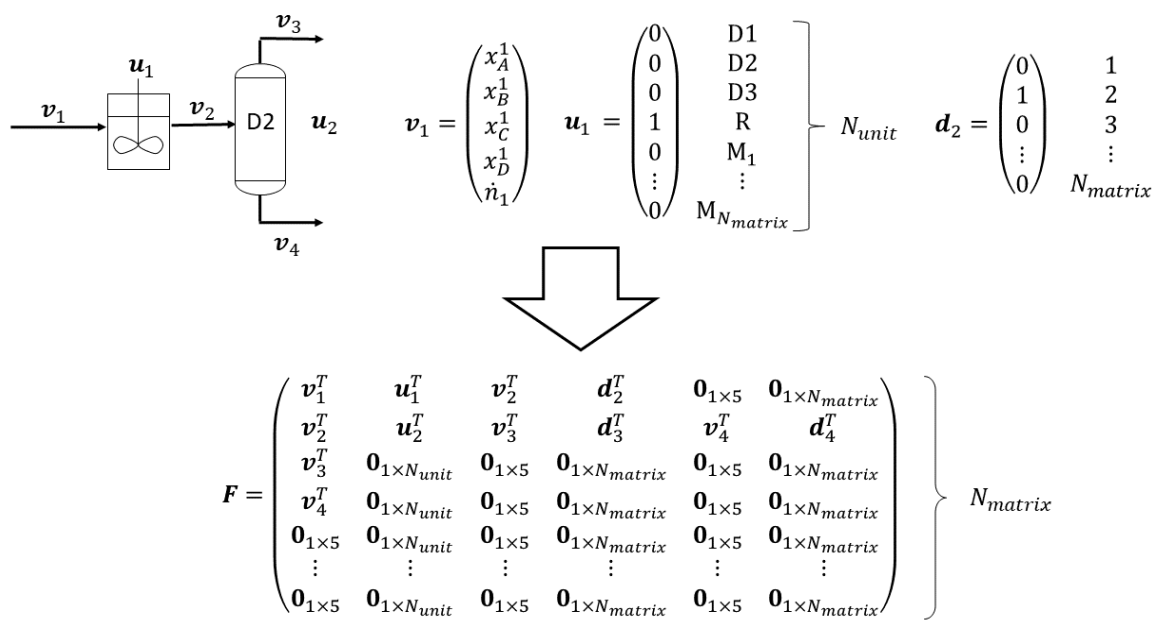

**Figure 2.** Construction of flowsheet matrix $\boldsymbol{F}$ along an example flowsheet. $\boldsymbol{F}$ contains the information of the stream table combined with structural information on the flowsheet. See text for an explanation of the nomenclature.

**Competitive two-player game.** The task of creating a profitable flowsheet is modelled as a two-player game. Each player obtains the same feed stream(s) and tries to create a more profitable flowsheet than the opponent does. The game is turn-based and at each turn, the active player selects one action and applies its flowsheet. Both players are always able to see their own and the opponents flowsheet. The game ends when both players have completed their flowsheets (either by using the termination action or if the flowsheet matrix $\boldsymbol{F}$ is full). The winner is the player with the flowsheet that has a larger net present value. If both players' flowsheets are equal in net present value, then the player that has completed the synthesis first wins the game. The winner obtains the reward $r = 1$, the loser $r = -1$. The agent is trained to master this game through self-play and RL. This means that the agent plays against itself and switches back and forth between the roles of player 1 and player 2 during the game. This setup enables using a modified version of the efficient training techniques to master the game of Go as proposed by Silver et al. 27., 28.. This will be explained in more detail below. After the training is complete and the agent has to solve a concrete problem, it will simply play another game against itself and select the winning flowsheet.

**Agent structure.** The agent consists of an ANN and a tree search. The ANN's outputs are used as inputs in the tree search that imitates a planning process. The structure of the ANN is depicted in Figure 3. The ANN receives the state vector $\boldsymbol{s}$ as input, which is constructed as concatenation of all rows of the flowsheet matrices $\boldsymbol{F_1}$ (current player) and $\boldsymbol{F_2}$ (waiting player). Further, a vector $\boldsymbol{g}$ is concatenated. It has the length of one row of $\boldsymbol{F_2}$ and flags whether the waiting player has already completed his/her flowsheet. $\boldsymbol{g}$ contains either only zeros (flowsheet of the waiting player is not completed) or only ones (flowsheet of the waiting player is completed). The ANN consists of one input layer, $N_{layer}$ fully connected hidden layers. Every hidden layer has $N_{node}$ nodes with ReLu-activation 29..

The ANN's output layer is fully connected and has two parts. On the one hand, there is a policy head that outputs the vector $\boldsymbol{\pi}$ of length $N_{action}$. Every entry corresponds to a probability with which the corresponding action should be executed by the agent. A softmax activation 30. is used to ensure that $\boldsymbol{\pi}$ has entries in the range [0,1] and sums up to 1. In the example of the present work, there are three distillation splits, one reactor and every stream can be mixed with any one of the remaining $N_{matrix} - 1$ ones. As the flowsheet can consist of up to $N_{matrix}$ open streams and one termination action is needed, there are $N_{action} = (N_{matrix}(4 + N_{matrix} - 1) + 1)$ theoretical actions. The policy might suggest infeasible actions, i.e. placing a unit at a stream that does not yet exist or is already connected to another unit. Therefore, the policy head $\pi$ is filtered. Entries that correspond to infeasible actions are set to 0. The remaining entries are scaled with a common factor, so that the resulting filtered vector $p$ also sums up to 1.

On the other hand, there is a value head that outputs a scalar $v$. Its output is generated using a tanh-activation 30. to ensure a value in the range $[1, -1]$. $v$ can be interpreted as an estimate of the reward at the end of the game for the current player.

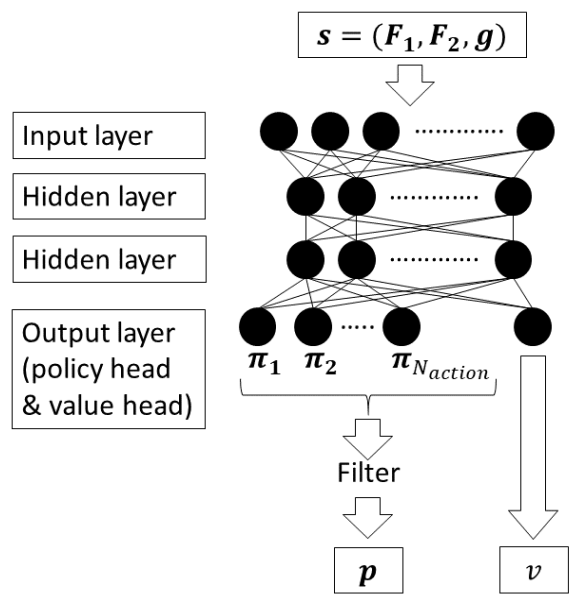

**Figure 3.** Structure of the agent's ANN in the SynGameZero method. The ANN has an actor-critic architecture. It calculates from the state input $s$ both a policy vector $\pi$ and a scalar value $v$. To obtain $p$, infeasible actions are filtered out of the vector $\pi$.

To improve its performance, the agent does not always select the action with the highest entry in $p$. Instead, $p$ and $v$ are used as the basis for a tree search to plan several actions in advance. The tree search imitates a typical human planning process that is made before finally deciding on an action. To avoid extensive computations, the tree search is adaptive in depth and does not use a full enumeration of all actions. Only promising actions are explored, where the values of $p$ and $v$ are used to quantify the word promising. The tree search is explained along Figure 4 and an example in which the agent (for the sake of simplicity) has only three possible actions named T, D1 and R, say terminating flowsheet synthesis, placing a distillation column on the first open stream, or placing a reactor on the first open stream. The tree consists of nodes and branches. The nodes $n \in \{I, II, III ...\}$ correspond to states of the two-player game. The branches $(n, a)$ correspond to the action $a$ that the current player takes at node $n$. The origin of the tree is the root node I. In Figure 4, the root node I corresponds to the beginning of the game when both players have an empty flowsheet. The current player is the one who takes the next action. His/her flowsheet is shown in the left half of the nodes. Since the game is turn-based, the order of the two flowsheets is switched after every action. Each node is either an explored node (e.g. node I in Figure 4) or an unexplored leaf node (e.g. node III). A node is explored if and only if the corresponding state vector $s$ is known. To explore an unexplored leaf node, i.e. to obtain its state vector $s$, the respective action has to be applied to the flowsheet of the current player in the parental node and the flowsheet simulation has to be evaluated. For example, if node III in Figure 4 has to be explored, then action D1 (adding a distillation column) has to applied to the left flowsheet in node I. The resulting flowsheet is evaluated by flowsheet simulation yielding the flowsheet matrix $F_2$ of the state of node III. The flowsheet matrix $F_1$ of the state in node III is equal to $F_2$ of the state in node I. Whenever a node is explored, it is checked whether it is terminal, i.e. a node in that both flowsheets are terminated (e.g. node V in Figure 4). The termination of a flowsheet may be caused either by the action T (terminate) or by a full flowsheet matrix. If at least one player has a non-terminated flowsheet (for example node VIII), the node is not terminal. In this case, the branches of all feasible actions and the corresponding (unexplored) leaf nodes are added to the tree below that node.

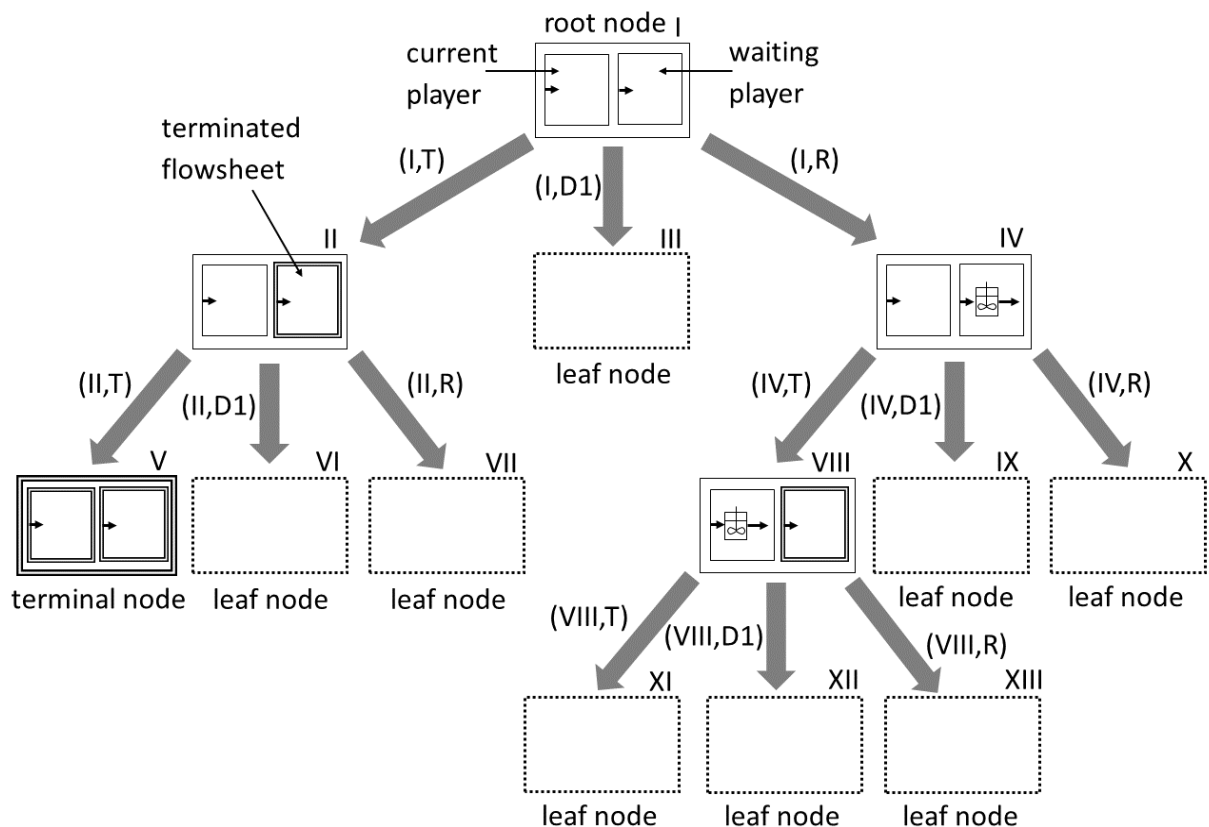

**Figure 4.** Example tree search at the beginning of the game (flowsheets of both players empty) with three possible actions {T, D1, R}. Unexplored leaf nodes are shown with dotted frames. Terminated flowsheets and terminal nodes are marked with bold frames. The order of the two flowsheets is switched after every action. The current player is the one who takes the next action. His/her flowsheet is shown in the left half of the nodes.

Four variables ($N_{n,a}, W_{n,a}, Q_{n,a}, P_{n,a}$) are stored for every branch ($n, a$) in the tree. The variable $N_{n,a}$ counts how often this branch has been taken during the tree search. The variable $W_{n,a}$ is the sum of all estimated and obtained rewards beneath that branch and $Q_{n,a}$ is defined as $W_{n,a}/N_{n,a}$. The values of $P_{n,a}$ are set to the corresponding value of the vector $\boldsymbol{p}$ that is obtained by feeding the state $\boldsymbol{s}$ of the node $n$ into the ANN. These variables are updated while the tree is constructed. They also guide both the extension of the tree and the final decision of the agent. A new tree is initialized only once at the beginning of every game. A root node with the state vector of two empty flowsheets is placed (cf. node I in Figure 4) and explored. The variables $N_{n,a}, W_{n,a}$ and $Q_{n,a}$ are set to 0 for the resulting branches, while the values of $P_{n,a}$ are obtained as described above. The tree search proceeds then in four steps:

**Step 1 Select.** The algorithm starts at the root node and runs down one path through the tree until it arrives at a leaf node or a terminal node $n_{bottom}$. At each node $n$, the algorithm greedily selects to follow the branch ($n, a$) that maximizes $Q_{n,a} + U_{n,a}(\alpha)$. $U_{n,a}(\alpha)$ is defined as follows:

$$U_{n,a}(\alpha) = P_{n,a} \frac{\sqrt{\sum_{b \in A} N_{n,b}}}{N_{n,a}+1} \qquad \text{if } Q_{n,a} > \alpha, \qquad (6)$$

$$U_{n,a}(\alpha) = 0 \qquad \text{if } Q_{n,a} \leq \alpha, \qquad (7)$$

where $A$ is the set of all actions. If there are two or more branches, that maximize $Q_{n,a} + U_{n,a}(\alpha)$, the branch among them with the largest value of $P_{n,a}$ is taken. To enhance exploration, the above greedy selection policy is replaced during training for the root node (and only the root node) by an $\varepsilon$-greedy policy 22.. With a probability of $\varepsilon$, an entirely random (i.e. uniform distribution) branch is selected. With a probability of $(1 - \varepsilon)$, the algorithm selects the branch using the above greedy policy that maximizes $Q_{n_{root},a} + U_{n_{root},a}(\alpha)$. In the present work, $\varepsilon$ is constant and equal to 0.2.

**Step 2 Explore and/or Evaluate.** If the node $n_{bottom}$ that was found in Step 1 is an unexplored leaf node, then it is explored and the resulting state vector $s$ is stored in $n_{bottom}$. Two cases might occur:

Case 1: The node $n_{bottom}$ is terminal. The winner of the game is determined by determining the net present value of both flowsheets. The reward is determined for the current player of the node $n_{bottom}$ and stored in the variable $V$ for Step 3.

Case 2: The node $n_{bottom}$ is not terminal. In this case, the state vector $s$ of the node $n_{bottom}$ is fed into the ANN. The value $v$ that is calculated by the ANN, is stored in the variable $V$ for Step 3.

**Step 3 Backup.** Starting at the node $n_{bottom}$ that was found in Step 1, the algorithm runs back upwards the tree until the root node. For every branch $(\tilde{n}, \tilde{a})$ passed on the way, the following updates are made to the branch variables:

$$N_{\tilde{n},\tilde{a}} = N_{\tilde{n},\tilde{a}} + 1, \tag{8}$$

$$W_{\tilde{n},\tilde{a}} = W_{\tilde{n},\tilde{a}} + tV, \tag{9}$$

$$Q_{\tilde{n},\tilde{a}} = W_{\tilde{n},\tilde{a}}/N_{\tilde{n},\tilde{a}}. \tag{10}$$

Herein, $V$ is the value that has been stored in Step 2. The variable $t$ takes into account that the agent switches players after every action when playing against itself. If $V$ has been determined at a node where player one is the current player, then $V$ is added ($t = 1$) at all branches that represent actions of player one. At the other branches, which represent actions of player two, $V$ is subtracted ($t = -1$). The opposite is done, when $V$ has been determined at a node where player two is the current player.

**Step 4 Play.** Steps 1-3 are repeated $K$ times as a loop, before the agent finally decides on an action at the root node of the tree. The decision is based on a probability vector $y$ with one entry for every feasible action at the root. The entry for action $a$ is calculated by

$$y_a = \frac{N_{n_{root},a}}{\sum_{b \in A} N_{n_{root},b}}. \tag{11}$$

During training, the decision is made by randomly selecting an action using the vector $y$ as probability distribution. After training, the moves are chosen greedily by always selecting the action $a_{play}$ with the largest $N_{n,a}$. If there are two or more branches that maximize $N_{n,a}$, then the algorithm selects the branch among them with the largest value for $P_{n,a}$. After the action is applied to the environment, the tree is shifted downwards. The node that is reached by the selected action $a_{play}$ becomes the new root node. The tree is cut off above. The values of the branch variables are retained.

The tree search algorithm is briefly interpreted in the following. The algorithm generally selects actions with large values of $N_{n,a}$. $N_{n,a}$ counts how often the branch has been taken during the selecting in Step 1. For success of the algorithm, it is therefore crucial to select promising actions in Step 1. This selection is based substantially on the value of $Q_{n,a} + U_{n,a}(\alpha)$. $Q_{n,a}$ is an estimate of the action value of action $a$ at node $n$. The action value is the state value $v$ of the game situation for the current player after he/she has selected the action $a$. The state value $v$ can only be determined with exact certainty at a terminal node. At other nodes, it is estimated by the ANN. In the backup (Step 3), the best available guess for $v$ is backed up along the search path to improve the estimation of $Q_{n,a}$. Using Equations (9) and (10), $Q_{n,a}$ is the average of the best available guesses for the state values $v$ of all explored nodes below the branch $(n, a)$. If the path selection in Step 1 would only depend on $Q_{n,a}$, then a wrong estimate of $Q_{n,a}$ at the beginning of the tree search might lead to inefficient exploration. Therefore, the function $U_{n,a}(\alpha)$ is also considered. This function is large for actions $a$ that have a large value $P_{n,a}$ but a small value $N_{n,a}$ compared to $\sum_{b \in A} N_{n,b}$. These actions are favored by the ANN, but have not been explored so far. If these promising actions would not be explicitly considered using $U_{n,a}(\alpha)$, they might be overlooked by chance at the beginning of the tree search. Later in the tree search, if such an action has turned out to be not constructive (the estimate $Q_{n,a}$ falls below the threshold $\alpha$ that is typically chosen rather low, e.g. -0.9), then the exploration function $U_{n,a}(\alpha)$ is no longer considered. This ensures that shortsighted recommendations of the ANN do not bias the tree search on the long run.

## 2.4 Training

The goal of the agent training is to adjust the parameters of the ANN, so that the ANN ideally predicts the consequences of a potential action up till the end of the game. The ideal ANN should output a value $v$ that correctly predicts the chances of the current player to win the game. The output $\boldsymbol{p}$ should be ideally a sharp distribution with a maximum at the action that maximizes the chances to win the game. The training procedure for approaching such behavior is outlined in the following. At the start of the training procedure, the ANN is initialized with random weights. In training, the agent plays a large number $N_{steps}$ of games against itself. The given feed stream(s) in the games can be varied randomly to obtain an agent that is able to solve a broad class of problems. For example, if an agent is desired that can separate a quaternary mixture for all possible feed compositions, then the feed compositions should be broadly sampled in the entire composition space during the training process. At the beginning of every game, the search tree is initialized with the given feed(s). Then the agent plays the game until the end (both players terminated their flowsheets). Thereby, every decision that had been made in Step 4 of the tree search is stored. Stored are the state vector $\boldsymbol{s}$ at the root and the vector $\boldsymbol{y}$ of the decision. After finishing the game, the decision data is augmented by the final reward $r$ that has been obtained at the end of game. The tuples of the form $(\boldsymbol{s}, \boldsymbol{y}, r)$ are stored in a memory of size $N_{memory}$. The oldest tuples are replaced in the memory, if the memory is full. After every game, a batch of $N_{batch}$ tuples is sampled randomly out of the memory. With this batch, the parameters of the ANN are optimized using stochastic gradient descent (SGD) 31.. Two optimization steps are performed. The first one with respect to the loss function $l_1$ and the second one with respect to the loss function $l_2$:

$$l_1 = (v - r)^2, \tag{12}$$

$$l_2 = \sum_{i=1}^{N_{action}} (\pi_i - y_i)^2. \tag{13}$$

## 2.5 Implementation

The flowsheet simulation, the agent, and the training were implemented with the programming language Python. For the ANN, the methods from the package Tensorflow (version 1.9.0) 30. were used. The SGD steps were performed using the Adam optimizer 30. with a learning rate $\beta$, cf. Table 2. The gradients are transformed by the function tensorflow.clip_by_global_norm 30. (with the argument 'clip_norm' equal to 5), which prevents the gradients of growing too large and causing instabilities during training.

## 3 Results and Discussion

### 3.1 Results

#### 3.1.1 Preliminary remarks

The results are shown along the quaternary system (A+B+C+D) and the process units described in the above subsection about the environment. Two case studies are done, which differ in the cost function and the number of feed streams. The cost function and number of feed streams were varied to obtain some variation in the results. The parameters for the cost functions are given in Table 1. The numerical tuning parameters of the algorithms are listed in Table 2.

**Table 1.** Investment costs $I_u$ for distillation D, reactor R and mixer M, and prices $p_i$ of compounds A, B, C, D used in the determination of the net present value in the present work.

|  | $I_D$ / k€ | $I_R$ / k€ | $I_M$ / k€ | $p_A$ / k€/kmol | $p_B$ / k€/kmol | $p_C$ / k€/kmol | $p_D$ / k€/kmol |
|---|---|---|---|---|---|---|---|
| Case 1 | 10000 | 10000 | 1000 | 1 | 1 | 1 | 1 |
| Case 2 | 10000 | 10000 | 1000 | -0.125 | -0.125 | 2 | 2 |

**Table 2.** Numerical tuning parameters used in the examples.

|        | $N_{steps}$ | $N_{matrix}$ | $N_{memory}$ | $N_{batch}$ | $N_{layer}$ | $N_{node}$ | $K$ | $\alpha$ | $\beta$ |
|--------|-------------|--------------|--------------|-------------|-------------|------------|-----|----------|---------|
| Case 1 | 5000        | 10           | 256          | 32          | 2           | 32         | 20  | -0.9     | 0.0001  |
| Case 2 | 20000       | 10           | 256          | 32          | 2           | 64         | 40  | -0.9     | 0.0001  |

The composition of the feed streams in the example problems are not fixed. Instead they are varied randomly to check the agent's performance over a wide range of feed compositions. To evaluate the success of the training the following procedure is used. One benchmark flowsheet is defined. This could be for example an educated guess by a human. Further, an evaluation set of 1000 problems is created. These problems differ in the composition of the feed stream(s) that are determined randomly. The trained agent synthesizes flowsheets for all 1000 problems. The flowsheets provided by the agent are compared to the respective benchmark flowsheets using the net present value. It is expected to design flowsheets that have the net present value of the benchmark flowsheets or a better one (the benchmark might not be the optimal flowsheet for all possible feed compositions). The success rate of the agent is defined using the following metrics.

$$R_1 = \frac{N_s}{1000} \tag{14}$$

$$R_2 = \frac{\sum_{j=1}^{1000} \frac{w_j}{b_j}}{1000} \tag{15}$$

$$R_3 = \frac{\sum_{j \in \Lambda} \frac{w_j}{b_j}}{|\Lambda|} \tag{16}$$

$$R_4 = \frac{\sum_{j \in \Gamma} \frac{w_j}{b_j}}{|\Gamma|} \tag{17}$$

$N_s$ is the number of problems for which the agent reached at least the net present value of the benchmark flowsheet. $R_1$ is the overall success rate. $w_j$ is the net present value of the agent's flowsheet and $b_j$ the net present value of the benchmark flowsheet in problem $j$, respectively. Thus, $R_2$ is the average of the agent's net present value relative to the benchmark. $\Lambda$ is the set of all problems, where the agent performed at least as good as the benchmark ($|\Lambda|$ is the number of elements in $\Lambda$). $\Gamma$ is the set of all other problems, i.e. where the agent performed worse than the benchmark ($|\Gamma|$ is the number of elements in $\Gamma$). Thus, $R_3$ and $R_4$ are similar to $R_2$ but they measure how well the agent succeeded or how bad it failed, respectively.

### 3.1.2 Case 1

In Case 1, there is one single feed stream and all components have the same prices. For most feed compositions, a reactor does not increase the value of stream and a distillation sequence should provide the most profitable process. During the training, random quaternary and ternary feed mixtures out of the following feed types were selected: $[x_A; x_B; x_C; x_D], [0; x_B; x_C; x_D], [x_A; 0; x_C; x_D], [x_A; x_B; 0; x_D], [x_A; x_B; x_C; 0]$. The mole fractions of the non-zero components were chosen randomly. The total molar flow rate of the feed was always set to 1 kmol/hr. Figure 5 shows examples (as training is a stochastic process) for the flowsheet that is proposed by the agent at different stages during training.

For evaluation of the trained agent, the benchmark flowsheet of three distillation columns as shown in the right panel of Figure 5 is defined. The training procedure was repeated 5 times. Each time, the resulting agent surpassed an overall success rate $R_1 = 0.98$.

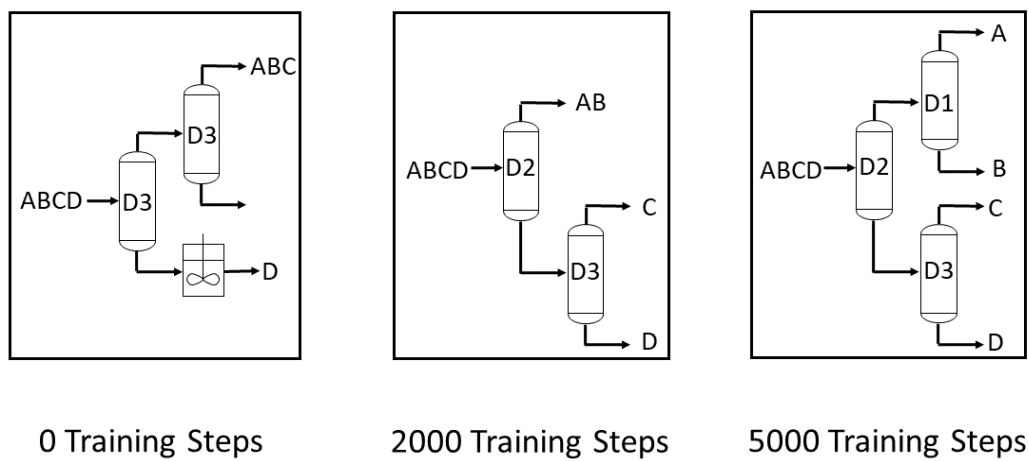

**Figure 5.** Illustrative example for the evolution of the agent during the training process in Case 1. Flowsheets proposed by the agent to separate an equimolar quaternary mixture are shown.

### 3.1.3 Case 2

In Case 2, the cost function is changed so that A and B have a negative price. C and D are high-value products. Thus, it is worth to react A and B to C and D. Four different feed situations are considered. Situation 1 considers two feed streams of the types: $[\dot{n}_A; \dot{n}_B; 0; 0]$, $[0; 0; \dot{n}_C; \dot{n}_D]$. Situation 2 considers two feed streams of the types: $[\dot{n}_A; 0; \dot{n}_C; 0], [0; \dot{n}_B; 0; \dot{n}_D]$. Situation 3 considers one feed stream of the type: $[\dot{n}_A; 0; \dot{n}_C; \dot{n}_D]$. Situation 4 considers one feed stream of the type: $[0; \dot{n}_B; \dot{n}_C; \dot{n}_D]$. For every game during training, one of the four situations is selected randomly. The given molar flowrates $\dot{n}_i$ are sampled randomly out of $[0.2, 1.2]\frac{\text{kmol}}{\text{hr}}$.

In Figure 6 we show examples (as training is a stochastic process) for the flowsheets that were proposed by the trained agent, depending on the feed situation. The panels 1), 2), 3) and 4) refer to feed situations 1, 2, 3 and 4, respectively.

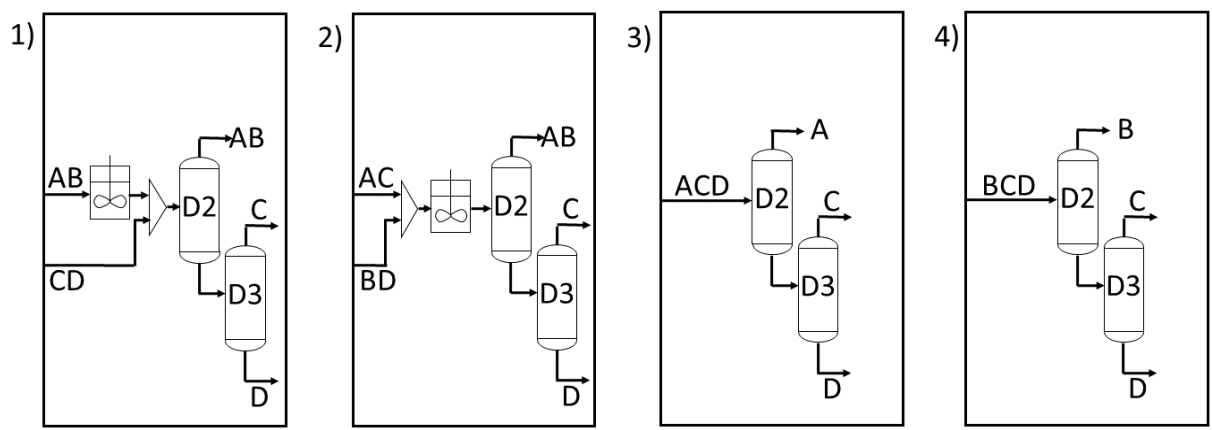

**Figure 6.** Example for the flowsheets proposed by the trained agent after the training process. The panels 1), 2), 3) and 4) refer to feed situations 1, 2, 3 and 4, respectively.

The evolution of the agent during the training process for situation 1 (feed streams: $[\dot{n}_A; \dot{n}_B; 0; 0], [0; 0; \dot{n}_C; \dot{n}_D]$) is shown in Figure 7. The rows correspond to different stages of training. The right panel shows an example for the flowsheet that was proposed by the agent at this stage. The left panel shows a 3D plot of three highlighted entries of the vector $\boldsymbol{p}$ (which depends on the ANN's output $\boldsymbol{\pi}$) at the beginning of the flowsheet synthesis, i.e. the ANN's suggestions for the very first action. The data is plotted over possible process feeds. Since the space of process feeds is 4-dimensional, we restrict us for the sake of illustration to a 2-dimensional subspace in which $\dot{n}_A = \dot{n}_B$ holds in the AB feed stream and $\dot{n}_C = \dot{n}_D$ holds in the CD feed stream. Action 1 is mixing both feed streams. Action 2 is placing a distillation column of type D3 at the CD feed stream. Action 3 refers to placing a reactor R at the AB feed stream.

At the beginning of the training, the actions of the agent are random. The suggested probabilities of the three highlighted actions are small as they are not significantly larger than the probabilities of any other feasible action (there are 10 feasible actions for the move). None of the shown actions is selected by the agent. Through feedback during training, the ANN favors (as first action) placing a distillation column that splits C and D at the CD feed stream (Action 2) after 2,000 training steps. After 6,000 training steps, the agent has learnt to complete the reaction part of the flowsheet before distillation is done. At the shown training example, the agent prioritizes mixing (Action 1) before reaction. At the end of training, the agent has learnt that bypassing the reactor with the products C and D yielding a higher conversion. C and D are separated only later together with the reactor outlet.

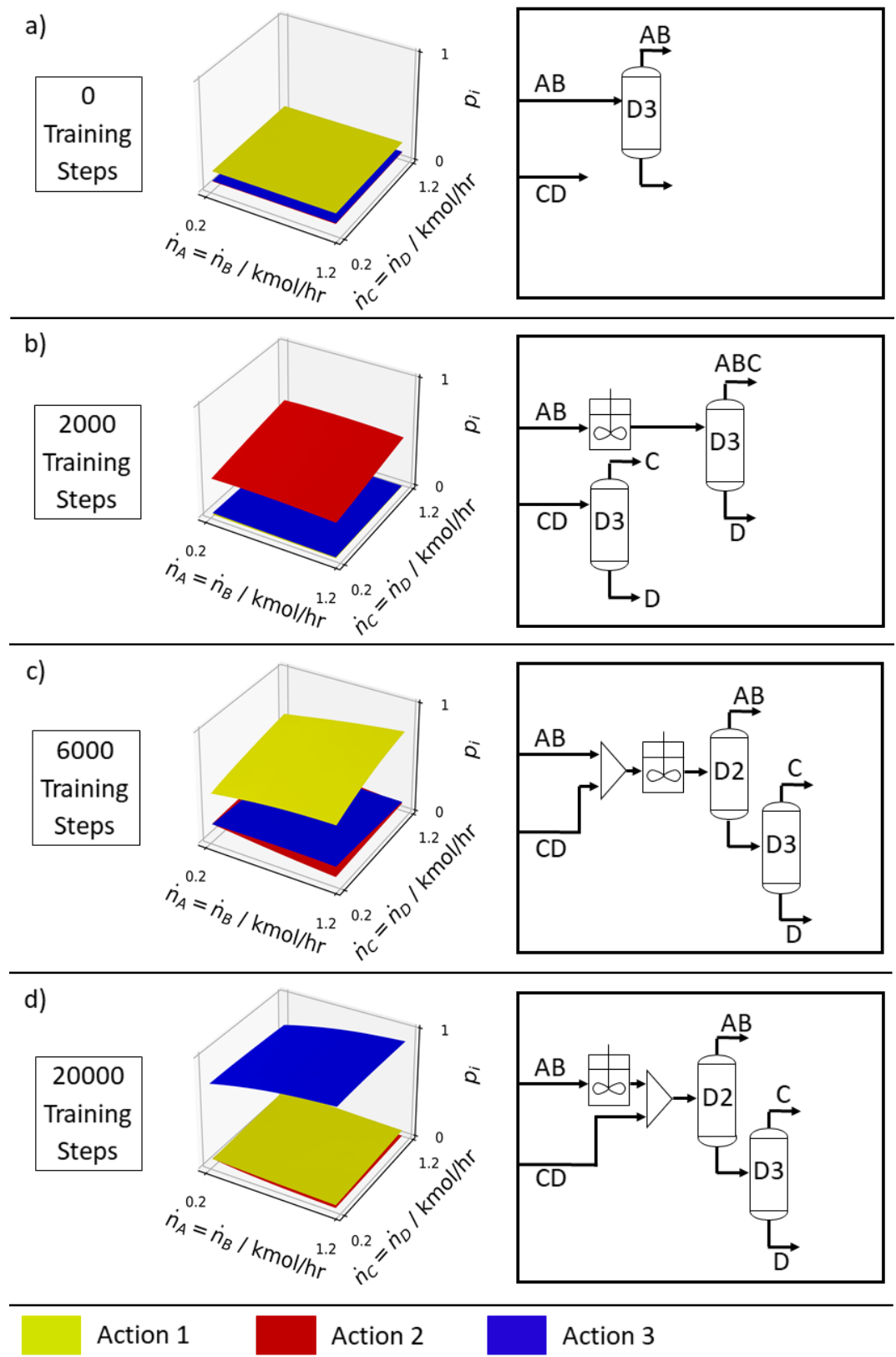

**Figure 7.** Example for the evolution of the agent during the training process for situation 1. The 3D plots show the value of three highlighted actions of the ANN's output vector $p$ over a subset of the feed space ($\dot{n}_A = \dot{n}_B$, $\dot{n}_C = \dot{n}_D$) for the first action of the agent. Action 1 is mixing both feed streams. Action 2 is placing a distillation column of type D3 at the CD feed stream. Action 3 refers to placing a reactor R at the AB feed stream.

The agent's performance is evaluated individually for all four feed situations. The flowsheets that are shown in Figure 6 are defined as the respective benchmark flowsheets. The training process was repeated 5 times. Table 3 shows the average of the performance metrics.

In situation 1, the agent meets or beats the benchmark in 84 % of the cases. In the successful cases, the possible gain over the benchmark is not large ($R_3 = 1.12$), indicating that the benchmark is already quite good. Although the agent is worse than the benchmark in 16 % of the cases, its average net present value is still slightly better than the benchmark's ($R_2 > 1$). In situation 2, the agent is almost always as good as the benchmark. At the cases, in which it is worse, it is only slightly worse. This indicates that there are several flowsheets with very similar net present values. Thus, the performance of the agent is still quite good. In the situations 3 and 4, there is either A or B missing in the process. Thus, there is no use for a reactor. The optimal flowsheets are distillation sequences for separating the ternary mixtures. The trained agent solves the problems in these situations without any difficulty.

**Table 3.** Average performance metrics as defined in Equations (14)-(17) for 5 training processes in Case 2. The metrics are evaluated individually for the feed situations 1-4. The corresponding benchmark flowsheets are shown in Figure 6.

|       | Situation 1 | Situation 2 | Situation 3 | Situation 4 |
|-------|-------------|-------------|-------------|-------------|
| $R_1$ | 0.84        | 0.99        | 1           | 1           |
| $R_2$ | 1.08        | 1.13        | 1           | 1           |
| $R_3$ | 1.12        | 1.13        | 1           | 1           |
| $R_4$ | 0.88        | 0.99        | -           | -           |

## 3.2 Discussion

The present work shows an RL-based approach to train an agent to solve basic flowsheet synthesis problems without the use of prior knowledge or heuristics. The agent consists of an ANN and a tree search in which the planning process is modelled within a two-player game. This setting allows the usage of a modified version of the training algorithm of Silver et al. 27., 28.. The trained agent mostly succeeds at the given problems, by combining discrete actions to synthesize a flowsheet using systematic generation. To assess the efficiency of the approach, the total number of possible flowsheets for a fixed feed composition has to be determined as follows for a matrix size of ($N_{matrix} = 10$). To the first stream, one could connect three distillation columns, one reactor or a mixer to one of the streams 2 to 9 (not to stream 10, as this would result in a new stream 11, which exceeds the size of the flowsheet matrix). Additionally it is possible to let the stream leave the process, which results in 13 possibilities. For the second stream one arrives at 12 possibilities (as mixing to stream 1 was already counted in beforehand). This calculation can be continued until stream 8 (6 possibilities). At stream 9 there is only the possibility of placing a reactor or letting that stream leave the process (distillation columns are not possible as the state matrix can only contain one more stream). At stream 10 no unit can be placed, therefore this stream leaves the process. This results into $2 \cdot 13!/5!$ possibilities, which is slightly above 100 million different flowsheets. This does not even count in that the feed compositions are sampled from a continuous range and therefore the state space for the agent is infinitely big (of course an engineer knows that sometimes different feed compositions can be approached by the same flowsheets, but the agent certainly has to learn this first). The tree size cannot exceed $2 \cdot K \cdot N_{matrix}$ (which is equal to 800 for Case 2). In the examples, the trees contained actually only an average of around 200-300 flowsheets in every training step, many of them visited several time from step to step. It is estimated that the number of total distinct flowsheets visited during the tree search is significantly smaller than 4 million. The ANN of the agent shows clear learning behavior, cf. Figure 7. Thus, it can be concluded that the winning flowsheets are not found by luck through massive enumeration in the tree search.

The process units and possible processes that were considered in the training are few in number and very basic in their modelling. This is enough for a proof-of-concept of the presented approach. However, the examples have several limitations that have to overcome in future work:

First, a larger number of process units and chemical compounds would require a larger flowsheet matrix. This would blow up both the agents input and the action space. The present work's representation of state and actions is mostly not scalable to very large problems. Alternative formulations and techniques might have to be used or developed. This may include feature extraction and convolutions for large inputs or the use of hierarchical agent decision structures like hierarchical neural networks 22..

Second, only processes without recycles were considered. Recycles are not a problem for the agent's structure or the RL method. They are just additional discrete action possibilities. Contrary, recycles are rather challenges for the simulation environment. Processes with recycles remain hard to simulate in automated fashion, although novel simulation methods let us be optimistic 32., 33.. Besides robustness, another challenge arises from recycles. If one would formulate the process units as we did in the present work (sharp splits in the distillation), then flowsheets with recycles would become infeasible or at least would make the simulation very stiff. Thus, the tasks of the process units have to be defined in a more tolerant way.

Another increase of complexity is the introduction of continuous parameters of the process units (e.g. pressures, temperatures, other specifications). This would lead to a hybrid discrete-continuous action space or at least to parametrized action spaces with discrete actions on a top level with subordinate continuous parameters. The algorithm of the present work only operates in a discrete action spaces and is not suited for these types of problems. There is already research on mixed action spaces in RL models 34.-36., however it is not straightforward to integrate these methods in our framework. Again, different agent structures or different ANN structures might be advisable for this type of problem.

The presented *SynGameZero* method that models the RL problem as a two-player game improved the efficiency of the RL significantly in the present work. It has two advantages: On the one hand, it eliminates the absolute value of the environment's native reward function (here: net present value). On the other hand, it has strong exploration abilities. Especially for player two who starts second, there is a strong motivation to explore novel actions instead of losing the game by just copying the actions of player one. This is because player two has the systematic disadvantage to lose the game if it is tied. During the training examples of the present work, player two lost the majority of all games as expected due to this disadvantage. However, it could be often observed that player two wins more games during the training phases is when new breakthrough actions are learnt. This indicates that these actions were found first by player two. Player one adopts these actions quickly by observing player two and benefits therefore as well. These beneficial features make the *SynGameZero* method certainly attractive also for other planning processes beyond chemical engineering.

## 4     Conclusion

A novel framework for automated synthesis of flowsheets was presented. The framework enables using RL for flowsheet synthesis. The physical knowledge of the chemical system and the process units is provided via process simulation. An agent without any prior knowledge or heuristics is successfully trained using RL to synthesize flowsheets by sequentially placing process units to an initially empty flowsheet. The net present value of the process is used as reward.

For the training of the agent, a novel RL method called *SynGameZero* was developed to efficiently master RL planning problems that have discrete action spaces and many local optima and, thus require good exploration schemes. The method models the RL problem as a turn-based game in that two-player compete to synthesize the best flowsheet. The method combines an ANN with classical actor-critical structure with an adaptive tree search for forward planning. Using relative instead of absolute rewards and systematic disadvantages of one of the players ensure good exploration abilities.

The framework and the *SynGameZero* method were tested in example synthesis problems. Therein, process simulations with four chemical compounds and simple short-cut models for the process units (reactor, distillation column and mixers) were evaluated. The agent's actions comprise placing a process unit or terminating the synthesis. The successfully trained agents of the examples demonstrate that both the overall framework for process synthesis by RL and the *SynGameZero* method work satisfactorily.

The presented approach for automated flowsheet synthesis is promising. In future work, extensions to more complex synthesis tasks should be studied. This includes, on the one hand, extensions of the process problems (e.g. larger processes, more process units, continuous parameters, recycles,…). On the other hand, improvements and extensions of the RL algorithm (e.g. hierarchical decisions, hybrid or parameterized action spaces, feature extraction) will be necessary to keep up with the process problems.

**References**

1. Westerberg A W. A retrospective on design and process synthesis. Comput. Chem. Eng., 2004, 28(4): 447-458

2. Stephanopoulos G, Reklaitis G V. Process systems engineering: From Solvay to modern bio- and nanotechnology. A history of development, successes and prospects for the future. Chem. Eng. Sci., 2011, 66(19): 4272-4306

3. Siirola J J. Strategic Process Synthesis: Advances in the Hierarchical Approach. Comput. Chem. Eng., 1996, 20(S2): S1637-S1643

4. Chen Q, Grossmann I E. Recent Developments and Challenges in Optimization-Based Process Synthesis. Annu. Rev. Chem. Biomol. Eng., 2017, 8: 249-83

5. Yeomans H, Grossmann I E. A systematic modeling framework of superstructure optimization in process synthesis. Comput. Chem. Eng., 1999, 23(6): 709-731

6. Stephanopoulos G, Westerberg A W. Studies in Process Synthesis II, Evolutionary Synthesis of Optimal Process Flowsheets. Chem. Eng. Sci., 1976, 31(3): 195-204

7. Zhang T, Sahinidis N V, Siirola J J. Pattern recognition in chemical process flowsheets. AIChE J., 2019, 65(2): 592-603

8. Gani R, O'Connell J P. A Knowledge Based System for the Selection of Thermodynamic Models. Comput. Chem. Eng., 1989, 13(4-5): 397-404

9. Kirkwood R L, Locke M H, Douglas J M. A prototype expert system for synthesizing chemical process flowsheets. Comput. Chem. Eng., 1988, 12(4): 329-343

10. Tula A K, Eden M R, Gani R. Process synthesis, design and analysis using a process-group contribution method. Comput. Chem. Eng., 2015, 81: 245-259

11. Daichendt M M, Grossmann I E. Integration of hierarchical decomposition and mathematical programming for the synthesis of process flowsheets. Comput. Chem. Eng., 1997, 22(1-2): 147-175

12. Martin M, Adams II T A. Challenges and future directions for process and product synthesis and design. Comput. Chem. Eng., 2019, 128: 421-436

13. Grossmann I E, Harjunkoski I. Process Systems Engineering: Academic and industrial perspectives. Comput. Chem. Eng., 2019, 126: 474-484

14. Stephanopoulos G. Artificial intelligence in process engineering - current state and future trends. Comput. Chem. Eng., 1990, 14(11): 1259-1270

15. Stephanopoulos G, Han C. Intelligent systems in process engineering: a review. Comput. Chem. Eng., 1996, 20(6-7): 143-191

16. Dimiduk D M, Holm E A, Niezgoda S R. Perspectives on the Impact of Machine Learning, Deep Learning, and Artificial Intelligence on Materials, Processes, and Structures Engineering. Integr. Mater. Manuf. Innov., 2018, 7: 157-172

17. Venkatasubramanian V. The promise of artificial intelligence in chemical engineering: Is it here, finally?, AIChE J., 2019, 65(2): 466-478

18. Lee J H, Shin J, Realff M J. Machine learning: Overview of the recent progresses and implications for the process systems engineering field. Comput. Chem. Eng., 2018, 114: 111-121

19. Eason J, Cremaschi S. Adaptive Sequential Sampling for Surrogate Model Generation with Artificial Neural Networks. Comput. Chem. Eng., 2014, 68: 220-232

20. Fahmi I, Cremaschi S. Process synthesis of biodiesel production plant using artificial neural networks as the surrogate models. Comput. Chem. Eng., 2012, 46: 105-123

21. Fernandes F A N. Optimization of Fischer-Tropsch Synthesis Using Neural Networks. Chem. Eng. Technol., 2006, 29(4): 449-453

22. Sutton R S, Barto A G. Reinforcement Learning: An Introduction. 2nd ed. Cambridge: The MIT Press, 2018

23. Lapan M. Deep Reinforcement Learning Hands-On. 1st ed. Birmingham: Packt Publishing Ltd., 2018

24. Shin J, Badgwell T A, Liu K-H, Lee J H. Reinforcement Learning - Overview of recent progress and implications for process control. Comput. Chem. Eng., 2019, 127: 282-294

25. Zhou Z, Li X, Zare R N. Optimizing Chemical Reactions with Deep Reinforcement Learning. ACS Cent. Sci., 2017, 3: 1337-1344

26. Khan A, Lapkin A. Searching for optimal process routes: A reinforcement learning approach. Comput. Chem. Eng., 2020, 141: 107027

27. Silver D, Schrittwieser J, Simonyan K, Antonoglou I, Huang A, Guez A, Hubert T, Baker L, Lai M, Bolton A, et al. Mastering the game of Go without human knowledge. Nature, 2017, 550: 354-359

28. Silver D, Hubert T, Schrittwieser J, Antonoglou I, Lai M, Guez A, Lanctot M, Sifre L, Kumaran D, Graepel T, et al. A general reinforcement learning algorithm that masters chess, shogi, and Go through self-play. Science, 2018, 362(6419): 1140-1144

29. Wang Y, Li Y, Song Y, Rong X. The Influence of the Activation Function in a Convolution Neural Network Model of Facial Expression Recognition. Appl. Sci., 2020, 10(5): 1897

30. Abadi M, Agarwal A, Barham P, Brevdo E, Chen Z, Citro C, Corrado G S, Davis A, Dean J, Devin M, et al. TensorFlow: Large-scale machine learning on heterogeneous systems. 2015, arXiv:1603.04467

31. Alpaydin E. Introduction to Machine Learning. 2nd ed. Cambridge: The MIT Press, 2010

32. Zinser A, Rihko-Struckmann L, Sundmacher K. Computationally Efficient Steady-State Process Simulation by Applying a Simultaneous Dynamic Method. Comput. Aided Chem. Eng., 2016, 38: 517-522

33. Hoffmann A, Bortz M, Burger J, Hasse H, Küfer K-H. A new scheme for process simulation by optimization: Distillation as an example. Comput. Aided Chem. Eng., 2016, 38: 205-210

34. Hausknecht M, Stone P. Deep reinforcement learning in parameterized action space. 2015, arXiv:1511.04143

35. Xiong J, Wang Q, Yang Z, Sun P, Han L, Zheng Y, Fu H, Zhang T, Liu J, Liu H. Parametrized deep q-networks learning: Reinforcement learning with discrete-continuous hybrid action space. 2018, arXiv:1810.06394

36. Neunert M, Abdolmaleki A, Wulfmeier M, Lampe T, Springenberg J T, Hafner R, Romano F, Buchli J, Heess N, Riedmiller M. Continuous-Discrete Reinforcement Learning for Hybrid Control in Robotics. 2020, arXiv:2001.00449v1